# A Quantum States Preparation Method Based on Difference-Driven Reinforcement Learning

Wenjie Liu*, Jing Xu† and Bosi Wang‡

*School of Software*
*Nanjing University of Information Science and Technology*
*Nanjing, Jiangsu 210044, P. R. China*
*\*wenjiel@163.com*
*†1156948326@qq.com*
*‡bosi@nuist.edu.cn*

Due to the large state space of the two-qubit system, and the adoption of ladder reward function in the existing quantum state preparation methods, the convergence speed is slow and it is difficult to prepare the desired target quantum state with high fidelity under limited conditions. To solve the above problems, a difference-driven reinforcement learning (RL) algorithm for quantum state preparation of two-qubit system is proposed by improving the reward function and action selection strategy. Firstly, a model is constructed for the problem of preparing quantum states of a two-qubit system, with restrictions on the type of quantum gates and the time for quantum state evolution. In the preparation process, a weighted differential dynamic reward function is designed to assist the algorithm quickly obtain the maximum expected cumulative reward. Then, an adaptive $\varepsilon$-greedy action selection strategy is adopted to achieve a balance between exploration and utilization to a certain extent, thereby improving the fidelity of the final quantum state. The simulation results show that the proposed algorithm can prepare quantum state with high fidelity under limited conditions. Compared with other algorithms, it has different degrees of improvement in convergence speed and fidelity of the final quantum state.

*Keywords*: Quantum state preparation; quantum system control; difference-driven reinforcement learning; two-qubit.

## 1. Introduction

The design of quantum system control is considered to be a necessary requirement for the establishment of powerful quantum information technology. The technology of accurately controlling quantum system through measurement, feedback and other methods[1,2] is the pillar of modern quantum computing, but the measurement process will destroy the superposition or entanglement state of quantum bits, making it to collapse into the result of measurement. At present, the demand for the application of quantum technology is growing, and it is a challenging task to realize the high fidelity control[3] of a single quantum system in the best way. Previous gradient-based control methods (including

‡Corresponding author.

SGD,[4] GRAPE[5] and variants[6]) are prone to fall into local optimization. The method based on genetic algorithm[7] requires a large amount of experimental data to optimize the control performance. In order to solve the above problems, reinforcement learning (RL)[8,9] has become an effective method to solve this series of complex problems, and has been successfully applied to various quantum tasks, including quantum state preparation,[8,10–14] quantum gate design[15] and quantum error correction.[16]

RL is a machine learning method without model feedback. It has been used in the learning control of quantum systems through the interaction between agents and their environment.[17,18] Chen *et al.*[19] proposed a probabilistic *Q*-learning method based on fidelity, which solved the balance between agent exploration and development, and applied it to the learning control of quantum systems. Bukov *et al.*[20] found that RL can effectively help find a short and high fidelity protocol in the nonintegrable multi-body quantum system of interacting qubits, from the initial state to the given target state. Niu *et al.*[21] proposed a general quantum control framework using deep RL g to optimize the speed and fidelity of quantum computing to prevent leakage and random control errors, and found a good quantum error correction strategy[22] to protect quantum bits from noise. In addition, Vedaie *et al.*[23] applied reinforcement learning algorithm to adaptive quantum enhanced metrology to realize multiphoton interferometry limited by single photon measurement. Cárdenas-López *et al.*[21] proposed a protocol for quantum reinforcement learning using quantum technology, which can be implemented in various quantum systems. Fösel *et al.*[24] showed how a network-based "agent" can discover a complete quantum error correction strategy to protect a group of qubits from noise. Yu *et al.*[25] used quantum RL to reconstruct photon qubit states to achieve maximum overlap. The above algorithm demonstrates from various angles that RL can help solve the control problem of quantum systems under a variety of quantum systems. However, for the two-qubit quantum system with limited control resources, due to the large state space of the two-qubit system, some quantum control methods use step reward functions, resulting in slow convergence speed, and it is difficult to prepare the required target quantum states with high fidelity under limited conditions. Therefore, there is still much room for improvement in such methods.

In order to solve the above problems, a quantum state preparation method based on difference-driven reinforcement learning (DIFF-RL) for two-qubit system is proposed. First, the task of quantum state preparation of two-qubit system is modeled to limit the type of quantum gate and the evolution of quantum state. In the preparation process, a weighted differential dynamic reward function is designed to help the algorithm quickly maximize the expected cumulative reward. Then, the adaptive $\varepsilon$-greedy action selection strategy is adopted to achieve the balance of exploration and utilization to a certain extent, so as to improve the fidelity of preparing the final quantum state in the two-qubit system. In order to verify the effectiveness and progressiveness of the algorithm, the quantum state preparation methods NN-QSC,[26] DRL-QSC[27] and ERL[28] based on RL are selected. The reward function and action selection strategy in the three algorithms are replaced with weighted differential dynamic reward function and adaptive $\varepsilon$-greedy action selection strategy, respectively, and compared with the original algorithm before replacement. Simulation results show that the proposed algorithm can prepare the desired target quantum states with high fidelity under limited conditions. Compared with other algorithms, it has different degrees of improvement in learning efficiency and fidelity of the final quantum states.

The main contributions of this paper include three aspects as follows:

(1) A weighted differential dynamic reward function is proposed to help the algorithm quickly obtain the maximum expected cumulative reward, so as to improve the convergence speed of RL algorithm;
(2) An adaptive $\varepsilon$-greedy action selection strategy is proposed to achieve the balance between exploration and utilization to a certain extent, so as to improve the fidelity of preparing the final quantum state in the two-qubit system;
(3) The simulation experiment is designed by using qiskit quantum computing framework and Linalg tool library. Compared with other methods, the effectiveness and progressiveness of the proposed algorithm are verified.

## 2. Elementary Knowledge

This section mainly introduces the basic knowledge of state control and RL of two-qubit quantum system.

## 2.1. *Two-quibit state*

A two-qubit state contain four ground states, which can be expressed as follows: $|00\rangle$, $|01\rangle$, $|10\rangle$ and $|11\rangle$. Because it can be in the superposition of these four ground states, the state of two qubits can be formulated as follows:

$$|\varphi\rangle = \alpha_{00}|00\rangle + \alpha_{01}|01\rangle + \alpha_{10}|10\rangle + \alpha_{11}|11\rangle, \quad (1)$$

the complex numbers $\alpha_{00}$, $\alpha_{01}$, $\alpha_{10}$ and $\alpha_{11}$ are called probability amplitudes, the measurement result of a two-qubit state can be $|00\rangle$, $|01\rangle$, $|10\rangle$ and $|11\rangle$, the probabilities are $|\alpha_{00}|^2$, $|\alpha_{01}|^2$, $|\alpha_{10}|^2$ and $|\alpha_{11}|^2$, satisfying $\sum_{x\in\{0,1\}^2}|\alpha_x|^2 = 1$, $|01\rangle$ represents a collection of symbol strings of size 2. Each element in the symbol string is an arbitrary combination of 0 and 1.

For the two-qubit system, one of the qubits can be measured separately. The results are shown in Table 1.

## 2.2. *State control of quantum systems*

For a finite dimensional closed quantum system, its state can be expressed by unit complex vector $|\psi\rangle$, and its dynamics can be described by Schrodinger equation

$$\frac{d}{dt}|\psi(t)\rangle = -\frac{i}{\hbar}\left(H_0 + \sum_{m=1}^{M} u_m(t)H_m\right)|\psi(t)\rangle, \quad (2)$$

where $\hbar$ is Planck constant, usually set to 1. $H_0$ is a free Hamiltonian[29–31] independent of time, and $H_m$ is a control Hamiltonian representing the interaction between the system and the external control field. The purpose of quantum state transformation is to use the control law $u^* = \{u_m^*(t)\}, t \in [0,T]$ to transform the quantum system from the initial state $|\psi_0\rangle$ to the target state $|\psi_f\rangle$. Fidelity is used to measure the similarity between the actual quantum state $|\psi(T)\rangle$ and the target quantum state $|\psi_f\rangle$, and further judge the performance of the control method.

In order to achieve the desired state, the gradient ascending pulse engineering (GRAPE) algorithm can be used to optimize the control field $u_m(t)$. For simplicity, time $T$ is usually discretized in $N_f$ steps, and the control field $u_m$ in each step is assumed to be constant. In order to calculate gradient information more conveniently, the cost function can be used as follows:

$$J(u) = |F(|\psi(T)\rangle, |\psi_f\rangle)|^2 = |\langle\psi(T)|\psi_f\rangle|^2. \quad (3)$$

For each control $u_m(t)$, the gradient direction of $\frac{\delta J(u)}{\delta u_m(t)}$ is as follows:

$$\frac{\delta J(u)}{\delta u_m(t)} = 2\mathrm{Re}[\langle\psi(T)|\psi_f\rangle$$
$$\langle\psi_f|U(T)U^\dagger(t)H_m U(t)|\psi_0\rangle], \quad (4)$$

where $U(t)$ meets $\frac{d}{dt}U(t) = -\frac{i}{\hbar}(H_0 + \sum_{m=1}^{M} u_m(t) H_m)U(t)$, $U(0) = I$, $\mathrm{Re}()$ is a function that takes the real part of a complex number. In order to maximize $J(u)$, the control field $u_m(t)$ is iteratively updated with the learning rate $\alpha$, and its formula is as follows:

$$u_m^{k+1}(t) = u_m^k(t) + \alpha\frac{\delta J(u)}{\delta u_m(t)}. \quad (5)$$

When the number of iterations reaches the maximum or a high fidelity close to 1, the learning process ends.

In the context of this experiment, the optional control set $\{u_j, j = 1, 2, \ldots, m\}$ of quantum system is given, in which each control $u_j$ corresponds to a unitary operation $U_j$. The goal of learning control is to learn a globally optimal control sequence $u^*$,

$$u^* = \underset{u}{\mathrm{argmax}}\, J(u). \quad (6)$$

Table 1. Two-qubit measurement results.

| Measuring position | Result | Probability | Status after measurement |
|---|---|---|---|
| First qubit | 0 | $|\alpha_{00}|^2 + |\alpha_{01}|^2$ | $|\varphi'\rangle = \frac{\alpha_{00}|00\rangle + \alpha_{01}|01\rangle}{\sqrt{|\alpha_{00}|^2+|\alpha_{01}|^2}}$ |
| First qubit | 1 | $|\alpha_{10}|^2 + |\alpha_{11}|^2$ | $|\varphi'\rangle = \frac{\alpha_{10}|10\rangle + \alpha_{11}|11\rangle}{\sqrt{|\alpha_{10}|^2+|\alpha_{11}|^2}}$ |
| Second qubit | 0 | $|\alpha_{00}|^2 + |\alpha_{10}|^2$ | $|\varphi'\rangle = \frac{\alpha_{00}|00\rangle + \alpha_{10}|10\rangle}{\sqrt{|\alpha_{00}|^2+|\alpha_{10}|^2}}$ |
| Second qubit | 1 | $|\alpha_{01}|^2 + |\alpha_{11}|^2$ | $|\varphi'\rangle = \frac{\alpha_{01}|01\rangle + \alpha_{11}|11\rangle}{\sqrt{|\alpha_{01}|^2+|\alpha_{11}|^2}}$ |

Table 2. Common symbols and descriptions of RL.

| Symbol | Describe |
| --- | --- |
| $s \in S$ | Finite state set, $s$ represents a specific state. |
| $a \in A$ | Finite action set, $a$ represents a specific action. |
| $T(S, a, S') \ldots P_r(s'\|s, a)$ | The transition model predicts the next state $s'$ according to the current state $s$ and action $a$, and $P_r$ represents the probability of taking action a from $s$ to transfer to $s'$. |
| $R(s, a) = \mathbb{E}[R_{t+1}\|s, a]$ | The instant reward after an agent takes an action. |

Aiming at a class of quantum control problems with limited control resources, this paper applies RL algorithm to quantum systems, and realizes the active control of quantum states by learning satisfactory control strategies.

## 2.3. Reinforcement learning

The main characteristic of RL is that it can interact with the environment of the agent.[32] Traditional RL optimizes numerical performance by learning a control strategy and making decisions in stages. Decision makers called agents interact with unknown environments in a trial and error manner, and occasionally receive feedback rewards that agents want to improve. This process is called Markov decision process. Table 2 provides several common symbols and their descriptions in RL.

The basic framework of RL is shown in Fig. 1, and the time step is expressed as $t \in [0, T]$, $T$ indicates the end time. Assuming that the state of the agent is $s_t$, it selects an action $a_t$ according to $a_t = \pi(s_t)$, in which the strategy function $\pi$ maps the state space to the action space. After executing action $a_t$, the agent will transfer from state $s_t$ to $s_{t+1}$ and obtain a scalar reward $r_t$. In order to achieve a trade-off between current incentives and future incentives, the discount factor is defined as $\gamma \in [0, 1]$, cumulative discount future rewards are defined as $R_t = \sum_{k=0}^{T-t} \gamma^k r_{t+k}$. Finally, the goal of training RL agent is to maximize $R_t$, in each state $s_t$, the best action $a^*$ can be determined.

## 3. Quantum State Preparation Based on Difference-Driven Reinforcement Learning Algorithm

### 3.1. Modeling of two-qubit quantum state preparation

The goal of quantum state preparation is to find a suitable control strategy. In a fixed total evolution time, the control quantum state evolves from the initial quantum state to the target state, and the final quantum state and the target quantum state have high fidelity.

The state vector of a two-qubit system can be formulated as follows:

$$|\psi\rangle = \sum_{ij} a_{ij} |i\rangle \otimes |j\rangle, \qquad (7)$$

where $|i\rangle$ and $|j\rangle$ represent the quantum ground states of the first and second qubits, respectively, $a_{ij}$ satisfies $\sum_{ij} |a_{ij}|^2 = 1$, and $\otimes$ represents the tensor product. The Hamiltonian of a two-qubit system can be expressed as follows:

$$H = [(S_z \otimes \mathbb{I})(\mathbb{I} \otimes S_z) + (S_y \otimes \mathbb{I} + \mathbb{I} \otimes S_y)] \\ + u_1(S_x \otimes \mathbb{I}) + u_2(\mathbb{I} \otimes S_x), \qquad (8)$$

where $H_0 = (S_z \otimes \mathbb{I})(\mathbb{I} \otimes S_z) + (S_y \otimes \mathbb{I} + \mathbb{I} \otimes S_y)$, $H_c = u_1(S_x \otimes \mathbb{I}) + u_2(\mathbb{I} \otimes S_x)$, $u_1$ and $u_2$ are external control fields acting in the direction of the first and second qubits, respectively. $S_x$, $S_y$ and $S_z$ are the spin operators which can be expressed as follows:

$$S_x = 0.5 \begin{pmatrix} 0 & 1 \\ 1 & 0 \end{pmatrix}, \quad S_y = 0.5 \begin{bmatrix} 0 & -i \\ i & 0 \end{bmatrix}, \\ S_z = 0.5 \begin{bmatrix} 1 & 0 \\ 0 & -1 \end{bmatrix}, \qquad (9)$$

$\mathbb{I}$ represents the identity matrix of size 2*2.

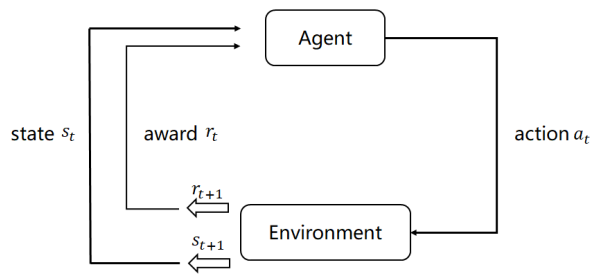

Fig. 1. Basic framework of RL.

Assuming that the quantum state at the current time is $|\psi_j\rangle$, the quantum state $|\psi_{j+1}\rangle$ at the next time can be expressed by the following time evolution equation:

$$|\psi_{j+1}\rangle = U_j|\psi_j\rangle, \quad (10)$$

$$U_j = e^{(-iH(u_j)dt)}, \quad (11)$$

where $U_j$ is the time evolution operator acting on the control field $u_j$ in time $[t_j, t_{j+1}]$, and $dt = t_{j+1} - t_j$ is the duration of the control pulse $u_j$. Suppose that for a fixed total evolution time $T$ and a specified number of control pulses $N$, the duration is $dt = \frac{T}{N}$.

Suppose that the fixed total evolution time $T \in [0, 4]$, the initial quantum state $|\psi_0\rangle = |01\rangle$, the target state $|\psi_T\rangle = |10\rangle$, and the control pulses $u_1$ and $u_2$ are limited by $u_1, u_2 \in \{2, 0, -2\}$. The fidelity of the following formula can be used to measure the state transition probability between the achieved final quantum state $|\psi_f\rangle$ and the ideal target quantum state $|\psi_T\rangle$,

$$F = |\langle \psi_f | \psi_T \rangle|^2, \quad (12)$$

where $\langle \psi_f |$ is the conjugate transpose of $|\psi_T\rangle$.

In order to quickly find a suitable control sequence $u$ to achieve high fidelity control tasks, difference-driven RL is proposed, which effectively improves the fidelity of the final quantum state.

### 3.2. *Difference-driven reinforcement learning*

Due to the large state space of the two-qubit system, some quantum control methods use the ladder reward function, which leads to slow convergence speed, and it is difficult to prepare the desired target quantum state with high fidelity under limited conditions. Different intermediate quantum states have different contributions to the subsequent evolution to the target quantum state, and the setting of rewards should also change accordingly. Here, a difference-driven RL algorithm is designed to solve the problem of two-qubit quantum state preparation. As shown in Fig. 2, a weighted difference dynamic reward function is designed to help the algorithm quickly maximize the expected cumulative reward and improve the convergence speed. Then, the adaptive $\varepsilon$-greedy action selection strategy is adopted to achieve the balance of exploration and utilization to a certain extent, so as to improve the fidelity of the final quantum state in the two-qubit system.

Inspired by the curiosity driven RL algorithm,[33] the complete reward is composed of internal rewards and external rewards. Usually, the internal rewards come from the gap between states, and the external rewards are set manually. Suppose an agent executes action $a_t$ when observing state $S_t$, and the environment returns to the next state $S_{t+1}$ and external reward $r_t^{\text{ext}}$. Since the external reward $r_t^{\text{ext}}$ can be sparse or nonexistent, if the internal reward $r_t^{\text{int}}$ is calculated by estimating the uncertainty in triple $(S_t, a_t, S_{t+1})$, the value function can be expressed as follows:

$$v_\pi(S) = \mathbb{E}_\pi \left[ \sum_{k=0}^{\infty} \gamma^k (r_{t+k+1}^{\text{ext}} + r_{t+k+1}^{\text{int}}) | S_t = S \right]. \quad (13)$$

With the progress of the training process, the internal reward will be reduced. When the internal

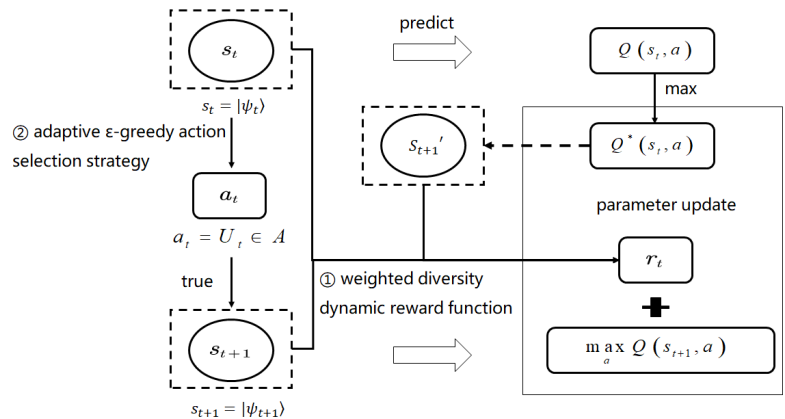

Fig. 2. Framework of difference-driven RL algorithm.

reward is small enough to be ignored, the value function $v_\pi(S)$ can be written as follows:

$$v_\pi(S) = \mathbb{E}_\pi\left[\sum_{k=0}^{\infty} \gamma^k r_{t+k+1}^{\text{ext}} | S_t = S\right]. \quad (14)$$

In other words, $v_\pi(S)$ will not be affected by internal rewards in the end. The intrinsic reward is estimated by considering the differences between the current quantum state, the real next quantum state and the predicted next quantum state and the target quantum state at the same time, and according to this difference, the possible reward for the agent to perform a quantum operation is dynamically set. Together with the adaptive $\varepsilon$-greedy action selection strategy, a quantum state preparation method based on DIFF-RL is formed.

### 3.2.1. *Design of weighted difference dynamic reward function*

Curiosity driven learning methods usually use internal rewards obtained under self-supervision to guide exploration in the environment. Figures 3(a) and 3(b) are two specific methods corresponding to such methods, namely Random Network Distillation[33] (RND) and Intrinsic Curiosity Module[34] (ICM).

RND proposes to quantify the novelty of states according to the distance between the output of a fixed and randomly initialized neural network and the next state predicted by another trained neural network. If the intrinsic reward of estimating the next state $S_{t+1}$ at time $t$ is called $r_t^N$, then, the state novelty in RND can be formulated as follows:

$$g(S_{t+1}; \theta_N) = h(S_{t+1}; \theta_N) - \phi(S_{t+1})_2^2. \quad (15)$$

ICM predicts the internal reward of the next state based on the current state and action of pairs. The loss between the predicted value and the real state is proportional to the final internal reward. If $r_t^F$ is the intrinsic reward for $S_t, a_t \to S_{t+1}$ path estimation, the state action novelty in ICM can be formulated as follows:

$$f(S_t, a_t, S_{t+1}; \theta_F) = m(\phi(S_t), a_t; \theta_F) - \phi(S_{t+1})_2^2. \quad (16)$$

Inspired by the above methods, a weighted difference dynamic reward function based on quantum state difference is proposed. Consider the difference between the current quantum state $S_t = |\psi_t\rangle$, the real next quantum state $S_{t+1} = |\psi_{t+1}\rangle$ and the predicted next quantum state $S'_{t+1} = |\psi'_{t+1}\rangle$ and the target quantum state $S_T = |\psi_T\rangle$ to estimate the internal reward. The real next quantum state is the next state that the agent reaches after actually executing a selected action $a_t \in U$, and the predicted next quantum state $S'_{t+1} = |\psi'_{t+1}\rangle$ is the next state that the neural network predicts after executing the action corresponding to the maximum $Q$ value that can be obtained under the current state. But it does not really enter the state of training in the next stage of RL. Therefore, it is hoped that the quantum state generated in the evolution of quantum state can approach the target quantum state and quickly evolve into the target quantum state. With the continuous training of RL, the difference between the quantum state and the target quantum state becomes smaller and smaller, but the fidelity $F$ is higher and higher. After the RL strategy agent executes an action, the closer it is to the target state, the greater the reward it should get. Fidelity $F$ can be calculated by formula (13), so the differences between the current quantum state, the real next quantum state and the predicted next quantum state and the target quantum state are formulated as follows:

$$d_t = F(S_t, S_T),$$
$$d_{t+1} = F(S_{t+1}, S_T), \qquad (17)$$
$$d'_{t+1} = F(S'_{t+1}, S_T).$$

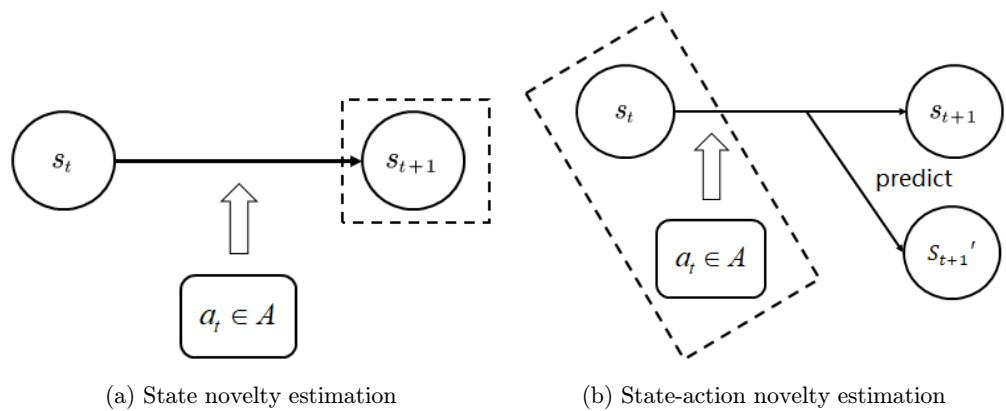

(a) State novelty estimation  (b) State-action novelty estimation

Fig. 3. Two different curiosity-driven RL algorithms.

Ideally, with the training, both $S_t = |\psi_t\rangle$, $S_{t+1} = |\psi_{t+1}\rangle$ and $S'_{t+1} = |\psi'_{t+1}\rangle$ will be closer to the target quantum state. $S_T = |\psi_T\rangle$, $d_t$ and $d_{t+1}$ will also become larger and larger, and the real-time reward $r_t$ that an agent can obtain in each state will also become larger and larger.

$$r_t = \frac{1}{3}d_t + \frac{1}{3}d_{t+1} + \frac{1}{3}d'_{t+1}. \qquad (18)$$

By giving different rewards to encourage agents to explore the environment, this method can quickly obtain a good RL strategy to accelerate the evolution of the control quantum system into the desired target state quantum state.

During each round of episode training, except that the initial state and target state are always consistent, other states under different learning rounds can be different, as shown in Fig. 4. In order to speed up the learning efficiency of RL and make the next learning round reach the final state closer to the target state in a fixed time, a weighted reward evaluation function is proposed.

Record all the state transitions in each round of training. During the current learning round of training, the fidelity $F_{i,j}$ between the intermediate quantum state $S_{i,j}$ and the target quantum state $S_T$ at the same position is compared with the previous round to obtain the quality $w_{i,j}$ of the training results of the previous and subsequent learning rounds,

$$w_{i,j} = \begin{cases} 1, & F_{i-1,j} < F_{i,j}, \\ 0, & F_{i-1,j} \geq F_{i,j}. \end{cases} \qquad (19)$$

If $F_{i-1,j} < F_{i,j}$, then it means that the fidelity between the $j$th intermediate quantum state and the target quantum state obtained in the $i-1$ learning round is greater than that between the $j$th intermediate quantum state and the target quantum state obtained in the $i-1$ learning round, and then set the current position of $w_{i,j}$ to 1, and each subsequent learning round of training is compared and assigned according to this rule, except that the initial state does not need to be compared, get a set of $w_{i,j}$.

In addition, we can not only calculate the good or bad degree of each real arriving state in each learning round by the above method, but also record the good or bad degree of the real next quantum state $S_{t+1} = |\psi_{t+1}\rangle$ and the predicted next quantum state $S'_{t+1} = |\psi'_{t+1}\rangle$ obtained by each state update in the training process of each learning round, which can be expressed as follows:

$$w_{i,j} = \begin{cases} 1, & F^t_{i-1,j} < F^t_{i,j}, \\ 0, & F^t_{i-1,j} \geq F^t_{i,j}, \end{cases}$$
$$y_{i,j} = \begin{cases} 1, & F^{t+1}_{i-1,j} < F^{t+1}_{i,j}, \\ 0, & F^{t+1}_{i-1,j} \geq F^{t+1}_{i,j}, \end{cases} \qquad (20)$$
$$z_{i,j} = \begin{cases} 1, & F^{(t+1)'}_{i-1,j} < F^{(t+1)'}_{i,j}, \\ 0, & F^{(t+1)'}_{i-1,j} \geq F^{(t+1)'}_{i,j}. \end{cases}$$

$w_{i,j}, y_{i,j}$ and $z_{i,j}$, respectively, represent the current quantum state $S_t = |\psi\rangle$ at each position, real next quantum state $S_{t+1} = |\psi\rangle$, and the predicted good or bad degree between the next quantum state $S'_{t+1} = |\psi'_{t+1}\rangle$ and the target quantum state $S_T = |\psi_T\rangle$.

For the $i$th learning round, it is also necessary to calculate the probability $W_{i-1}, Y_{i-1}, Z_{i-1}$ that $w_{i-1,j}, y_{i-1,j}, z_{i-1,j}$ are 1 in the training of the previous learning round,

$$W_{i-1} = \sum_{j=1}^{N} \frac{w_{i,j}}{N},$$
$$Y_{i-1} = \sum_{j=1}^{N} \frac{y_{i,j}}{N}, \qquad (21)$$
$$Z_{i-1} = \sum_{j=1}^{N} \frac{z_{i,j}}{N},$$

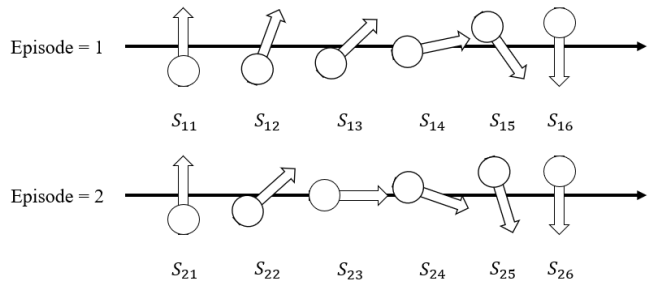

Fig. 4. Schematic of state update in RL under different episodes.

where $N$ represents the maximum number of iterations of training in each learning round, or the number of states obtained. Take the obtained probability of good or bad degree of different angles of the previous learning round as the weight of the current quantum state, the real next quantum state and the difference between the predicted next quantum state and the target quantum state in Eq. (18), then Eq. (18) can be rewritten as follows:

$$r_t = \begin{cases} \dfrac{1}{3}d_t + \dfrac{1}{3}d_{t+1} + \dfrac{1}{3}d'_{t+1}, & i = 1, \\ W_{i-1}d_t + Y_{i-1}d_{t+1} + Z_{i-1}d'_{t+1}, & i > 1. \end{cases} \quad (22)$$

In addition to using one-third of the training in the first learning round to represent the weight, in other cases, the reward can not only be dynamically updated according to the training of each single learning round, but also use the training of the previous learning round as the basis to help the RL algorithm quickly get the desired target state, i.e., the quantum state can evolve from the initial quantum state to the target quantum state with high fidelity and efficiency.

### 3.2.2. *Adaptive ε-greedy action selection strategy*

How to balance the relationship between exploration and utilization is a problem that must be considered in quantum control methods based on reinforcement learning. Exploration means that when selecting actions, the agent performs unknown $w_{i,j}$ or not optimal actions in order to expand the scope of state search. Exploring more unknown actions may not be able to perform the optimal action, and using the existing strategies to select the optimal action may not be able to explore the optimal control strategy.

The existing action selection strategies mainly include $\varepsilon$-greedy strategy and Boltzmann strategy. In strategy $\varepsilon$-greedy, each time the agent selects an action, it randomly selects the action to be executed with a probability of $\varepsilon$, and selects the action with the highest $Q$ value in the current state with a probability of $1 - \varepsilon$. Boltzmann strategy adopts the form of Boltzmann distribution, and the probability of selecting actions can be formulated as follows:

$$p(a_i|s) = \frac{e^{Q(s,a_i)/T}}{\sum_{k=1}^{N} e^{Q(s,a_k)/T}}, \quad (23)$$

where $p(a_i|s)$ represents the probability that the agent selects an action $a_i$ from the action set in the case of state $s$, $T$ represents the randomness of the selected action, and the greater the value of $T$, the higher the randomness. Since both methods need to choose appropriate strategies according to the specific application fields, and the setting of relevant parameters is basically based on experience, it is very difficult to determine a perfect parameter. The existing three methods of quantum state preparation based on reinforcement learning NN-QSC, DRL-QSC and ERL algorithms are based on the $\varepsilon$-greedy strategy. According to the change of the number of learning rounds, the number of learning rounds per fixed value is subtracted from the preset fixed value to reduce the size of $\varepsilon$, so that the action selection of the algorithm changes from exploration to utilization as the training proceeds. However, researchers with sufficient experience in this method have set relevant parameters according to a large number of experiments.

In order to adaptively change the value of $\varepsilon$ in the preparation process of quantum states according to RL, so as to help balance the relationship between exploration and utilization, an adaptive $\varepsilon$-greedy action selection strategy is proposed. By introducing the $\lambda$ parameter, the fidelity between the quantum state and the target quantum state in the quantum state control process is recorded more than 0.8 times. As the training proceeds, $\varepsilon$ is updated with the following calculation method:

$$\varepsilon = \varepsilon_{\min} + (\varepsilon_{\max} - \varepsilon_{\min})e^{-\frac{\lambda}{C}}, \quad (24)$$

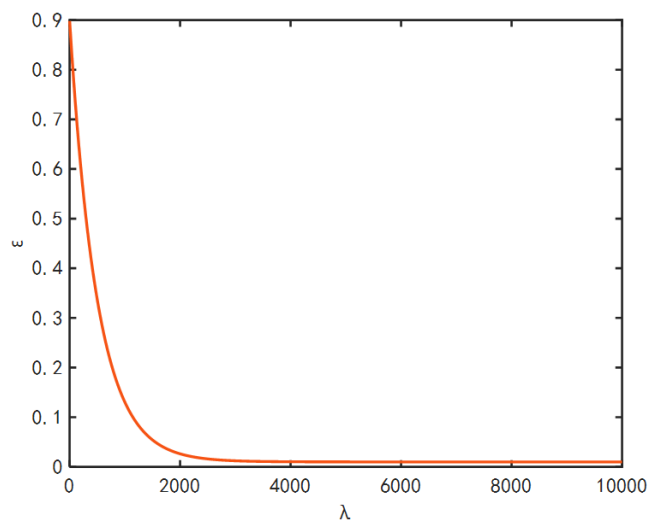

Fig. 5. Curve of adaptive $\varepsilon$-greedy calculation formula obtained by MATLAB.

**Algorithm 1.** Quantum state preparation method based on difference-driven reinforcement learning

---

Input: initial quantum state $|\psi_0\rangle$, target quantum state $|\psi_F\rangle$, quantum operation set $U$, fidelity $F$, discount factor $\gamma$, memory size $M$, batch size $B$, maximum learning round $K$, $\varepsilon_{\min} = 0.01$, $\varepsilon_{\max} = 0.9$, $C = 500$, $\lambda = 0$;

Output: optimal control parameters $x^* = [x^1, x^2, \ldots]$, where $x^t = \operatorname*{argmax}_{a}(Q(s_t, a; \xi))$;

Initialize evaluation network and target network parameters $\xi_1$, $\xi_2$, evolutionary experience pool $Me = \{E_1, E_2, \ldots, E_m\}$;

**for** $k = 1, 2, \ldots, K$ **do**

Initialization $t = 1$, $S_t = |\psi_t\rangle$;

   **for** $t = 1, 2, \ldots, T$ **do**

      Select action $a_t \in U$ and execute it under the adaptive greedy action selection strategy to obtain the next state $S_{t+1} = |\psi_{t+1}\rangle$ and the predicted next quantum state $S'_{t+1} = |\psi'_{t+1}\rangle$ obtained by executing the predicted $a'_t = \operatorname*{argmax}_{a}\{Q(s, a, \xi_1)\}$;

      Calculate $w_{i,j}$, $y_{i,j}$, $y_{i,j}$, $z_{i,j}$, $W_{i-1}$, $Y_{i-1}$ and $Z_{i-1}$ respectively according to formula (21) and formula (22), Using equation (23) to get weighted difference rewards $r_t$;

      Calculate timing difference error $\delta_t$ and store experience $E_t = (s_t, a_t, r_t, s_{t+1})$ in experience pool;

      Extract batch $B$ sample data $\{e_i\}_{i=1}^{B}$ from $Me$, according to $p_t = \frac{\delta_t^{\lambda}}{\sum_k \delta_t^{\lambda}}$, updating parameters $\xi_1$;

      Loop traversal, until $t = T$;

   **end for**

   Update parameters $\xi_2 = \xi_1$;

**end**

---

among them, $\varepsilon_{\min}$ and $\varepsilon_{\max}$ represent the minimum and maximum allowable $\varepsilon$ sizes, respectively, which are set to 0.9 and 0.01 here, and $C$ represents the update cardinality, which is set according to different tasks, which is set to 500 here. The function image is made by MATLAB, as shown in Fig. 5. The value of $\varepsilon$ decreases adaptively with the change of $\lambda$ value. As the training continues, the action selection changes from exploration to utilization. When the algorithm converges, the agent selects the best quantum operation to act on the quantum bit according to the optimal strategy obtained after convergence, helping the evolution of quantum states.

To sum up, taking ERL algorithm as the basic framework, this paper introduces the quantum state preparation method based on difference-driven RL, and gives the following algorithm process description.

## 4. Simulation Experiment

### 4.1. *Experimental design*

Because it is difficult to verify the effectiveness and efficiency of the algorithm on a real quantum computer, the algorithm is verified in the form of simulation experiments, using qiskit quantum computing framework and Linalg tool library. All codes are implemented on a single CPU+GPU workstation (CPU:Intel Xeon Gold 5218, GPU: GeForce RTX 2080 Ti 11G). The proposed DIFF-RL algorithm is applied to solve the problem of two-qubit quantum state preparation in a fixed time. In order to verify the effectiveness and progressiveness of the algorithm, this experiment chooses to use the quantum state preparation methods NN-QSC[26] and DRL-QSC[27] based on RL and the enhanced reinforcement learning (ERL) algorithm[28] to replace the reward function in the three algorithms with the dynamic reward function in DIFF-RL, and change the action selection strategy and compare it with the original algorithm before replacement to verify whether the proposed DIFF-RL algorithm can effectively help to efficiently and stably control the preparation of the required quantum state in the two-qubit scenario, so that the quantum state can quickly and efficiently evolve from the initial state to the required target state. The parameters and corresponding parameter values required for the simulation experiment are given in Table 3.

### 4.2. *Experimental results and analysis*

In order to verify whether the weighted reward evaluation function can effectively improve the

Table 3. Parameter settings used in simulation experiment.

| Parameter name | Parameter value |
| --- | --- |
| Learning rate | 0.005 |
| Incentive discount factor | 0.99 |
| Maximum learning rounds | 500 |
| Experience pool size | 2000 |
| Minimum batch times | 64 |
| Number of control pulses $N$ | 5, 10, 20, 30 |

*W. Liu, J. Xu & B. Wang*

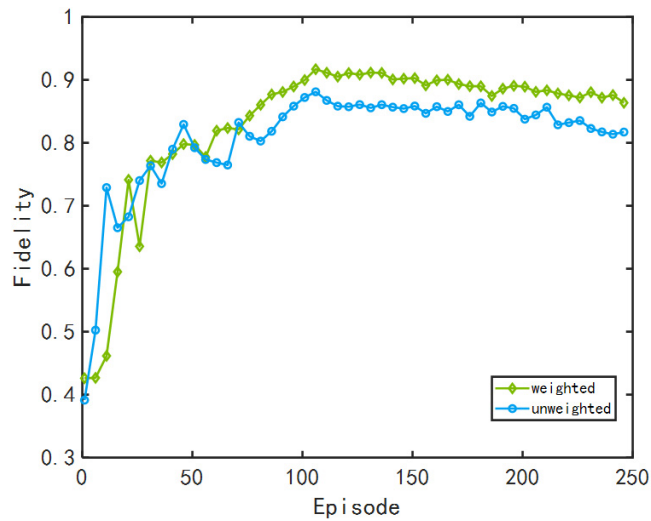

Fig. 6. Comparison of fidelity results obtained by pre-weighted and post-weighted algorithms.

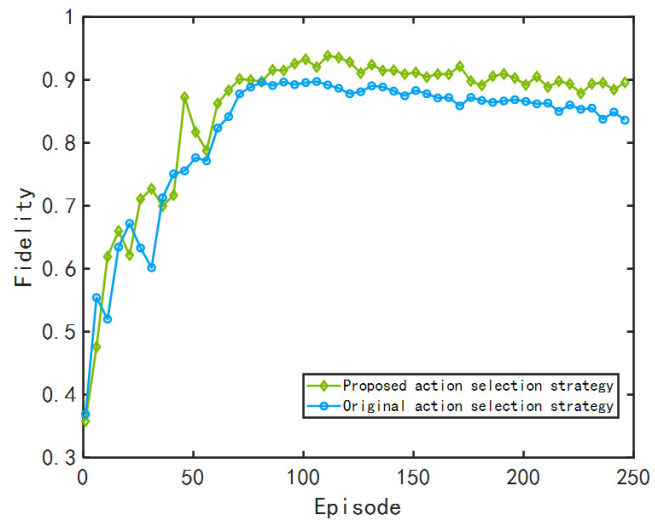

Fig. 7. The fidelity comparison results obtained by using adaptive action selection strategy and original strategy.

evolution results of quantum states, the experiment uses ERL algorithm as the basic algorithm, changes the reward function, directly uses the non-dynamically weighted form as shown in formula (18), and uses the weighted reward function as shown in formula (22), and calculates the fidelity between the final quantum state and the target quantum state after each learning round when the number of pulses is controlled $N = 10$, the results are shown in Fig. 6.

In order to verify the influence of adaptive $\varepsilon$-greedy action selection strategy on quantum control, this experiment replaces the original $\varepsilon$-greedy action selection strategy on the basis of ERL algorithm. Set the initial $\varepsilon$ of the original $\varepsilon$-greedy action selection strategy to 0.91, reduce the end of each learning round by 0.005, and set the minimum $\varepsilon$ to 0.01. When controlling the number of pulses $N = 10$, calculate the fidelity between the final quantum state and the target quantum state obtained after each learning round. The results are shown in Fig. 7. It can be seen from the figure that compared with another strategy, the adaptive $\varepsilon$-greedy action selection strategy not only improves the convergence speed, but also improves the fidelity of the final quantum state and the target quantum state.

Table 4 shows the average fidelity results of 50 tests using NN-QSC, DRL-QSC, ERL algorithms and the proposed DIFF-RL algorithm under different control steps. It can be seen from Table 4 that with the increase of the number of control pulses, the RL algorithm can perform more actions during the training process. When the training is over, the fidelity of the final quantum state obtained is gradually improved compared with the target quantum state. When the number of control pulses is certain, the results obtained by different algorithms are quite different. When $N = 5$, NN-QSC and DRL-QSC algorithms may fall into local optimization or not fully converge, and the average fidelity obtained is far less than that of ERL algorithm. With the increase of $N$, the algorithm can show better performance. When the three selected algorithms used as the benchmark are combined with the proposed DIFF-RL algorithm, the efficiency of quantum state evolution has been significantly improved, especially ERL+DIFF-RL algorithm, which not only has a better quantum state preparation effect under the condition of small number of control pulses, but also with the increase of $N$, the fidelity of the final quantum state and the target quantum state is close to 0.999.

Table 4. Average fidelity results of different algorithms under different numbers of control pulses.

| Number of control pulses $N$ | Algorithm name | | | | | |
| --- | --- | --- | --- | --- | --- | --- |
| | NN-QSC | DRL-QSC | ERL | NN-QSC+DIFF-RL | DRL-QSC+DIFF-RL | ERL+DIFF-RL |
| 5 | 0.8350 | 0.8839 | 0.9361 | 0.8333 | **0.9393** | **0.9513** |
| 10 | 0.8622 | 0.8750 | 0.9282 | **0.8840** | **0.9592** | **0.9679** |
| 20 | 0.8638 | 0.9423 | 0.9744 | **0.9051** | **0.9772** | **0.9917** |
| 30 | 0.9056 | 0.9747 | 0.9767 | **0.9274** | **0.9790** | **0.9936** |

Figure 8 shows the fidelity comparison results between the final quantum state and the target quantum state obtained at the end of each learning round in the training process of the six algorithms when the number of control pulses $N = 20$. Among them, algorithms 1–4 represent ERL algorithm, DRL-QSC algorithm, NN-QSC algorithm and DIFF-RL algorithm proposed in this paper. It can be seen from the figure that in the first few rounds of learning, the fidelity results obtained by all algorithms change violently, indicating that the RL algorithm has not fully converged. With the continuous training, the algorithm can find a suitable quantum state preparation strategy, so that the algorithm can control the conversion of quantum states to target states in a two-qubit closed quantum system with limited control resources, but the performance of different algorithms is different. Therefore, the gap between the final quantum state and the target state that can be obtained after the final convergence is also different.

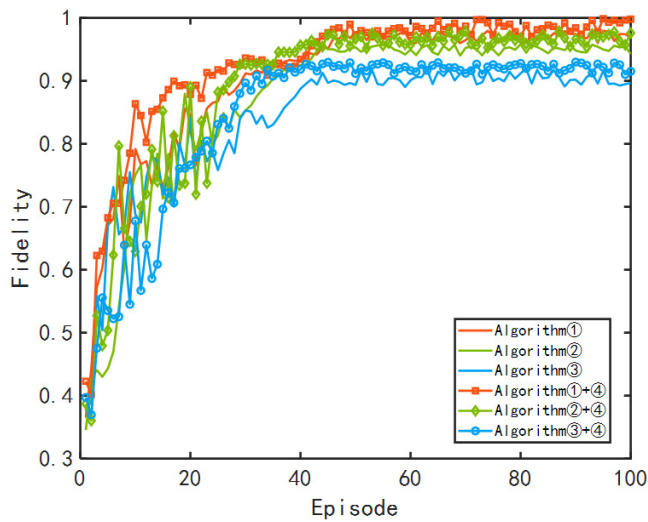

Fig. 8. Fidelity comparison results of different algorithms.

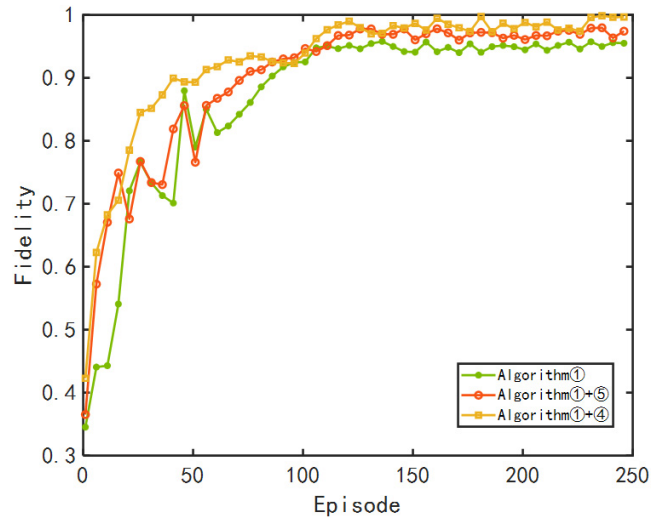

Fig. 9. Fidelity comparison results of three-reward function setting methods.

In addition, Fig. 9 shows the fidelity comparison results obtained when the number of pulses $N = 30$ is controlled by using the ERL algorithm with fixed reward and replacing the two schemes with dynamic setting reward function, respectively. Bukov et al.[10] selected the scheme of dynamically setting the reward function as shown below to verify the progressiveness of this method:

$$r_t = \begin{cases} 0, & t < T, \\ F(T) = |\langle \varphi | \varphi_T \rangle|^2, & t = T. \end{cases} \quad (25)$$

For the convenience of description, set it as 5 in the legend. It can be seen from Fig. 9 that the way of dynamically setting the reward function can effectively improve the learning process of quantum state preparation strategy and accelerate the rapid convergence of RL algorithm. Compared with algorithm 5, the final quantum state obtained by the proposed difference-driven RL algorithm has a relatively large improvement in convergence speed and final fidelity.

## 5. Conclusion

In this paper, a quantum state preparation method based on DIFF-RL is proposed. Weighted differential dynamic reward function and adaptive $\varepsilon$-greedy action selection strategy are proposed, respectively, which improves the convergence speed of the algorithm and fidelity of the final quantum state.

This paper only focuses on the time-dependent two-qubit closed quantum system. As we all know, high-dimensional (i.e., multi-qubit) quantum systems or open quantum systems are easily affected by the external environment, and they have more complex quantum characteristics. On the one hand, the follow-up work can study the control methods of a class of complex quantum subsystems. On the other hand, we will continue to explore more efficient reinforcement learning algorithms, such as multi-agent reinforcement learning, and study the continuous control tasks in quantum systems.

## Acknowledgments

This work is supported by the National Natural Science Foundation of China (62071240), the Innovation Program for Quantum Science and Technology (2021ZD0302902), and the Priority Academic Program Development of Jiangsu Higher Education Institutions (PAPD).